\documentclass[11pt]{article}
\usepackage{epsfig}
\usepackage{graphics}

\begin{document}
\centerline{\large\bf{Confirmation of the Luminous Blue Variable
status of MWC\,930}}

\vspace*{0.5cm}

\noindent {\bf A.~S.~Miroshnichenko$^1$, N.~Manset$^2$,
S.~V.~Zharikov$^3$, J.~Zsarg\'o$^4$, J.~A.~Ju\'arez Jim\'enez$^4$,
J.~H.~Groh$^5$, H.~Levato$^6$, M.~Grosso$^6$, R.~J.~Rudy$^7$,
E.~A.~Laag$^7$, K.~B.~Crawford$^7$, R.~C.~Puetter$^8$, D.~E.
Reichart$^9$, K.~M.~Ivarsen$^9$, J.~B.~Haislip$^9$,
M.~C.~Nysewander$^9$, A.~P.~LaCluyze$^9$}

\vspace*{0.5cm}

{\small \noindent $^1$Department of Physics and Astronomy,
University of
North Carolina at Greensboro, Greensboro, NC 27402, USA\\
$^2$CFHT Corporation, 65--1238 Mamalahoa Highway, Kamuela, HI 96743\\
$^3$Instituto de Astronom\'ia, Universidad Nacional Aut\'onoma de
M\'exico, Apartado Postal 877, 22830, Ensenada, Baja California,
M\'exico\\
$^4$Departimento de F\'isika, Escuela Superior de F\'isica y
Matematicas del Instituto Politecnico Nacional, Edif. 9,
U.~P.~Zacatenco, 07738, M\'exico D.~F., M\'exico,\\
$^5$Geneva Observatory, Geneva University, Chemin des Maillettes 51,
CH--1290 Sauverny, Switzerland\\
$^6$Instituto de Ciencias Astron\'omicas de la Tierra y del Espacio,
CONICET, Casilla de Correo 49, CP 5400, San Juan, Argentina\\
$^7$The Aerospace Corp., M2/266, P.O.Box 92957, Los Angeles, CA
90009, USA\\
$^8$Center for Astrophysics and Space Science, University of
California at San Diego, 9500 Gilman Drive, La
Jolla, CA 92093, USA\\
$^9$Department of Physics and Astronomy, University of North
Carolina, Chapel Hill, NC 27599, USA}

\vskip 0.5cm

{\bf Abstract.} We present spectroscopic and photometric
observations of the emission-line star MWC\,930 (V446\,Sct) during
its long-term optical brightening in 2006--2013. Based on our
earlier data we suggested that the object has features found in
Luminous Blue Variables (LBV), such as a high luminosity ($\sim
3\,10^{5}$ L$_{\odot}$), a low wind terminal velocity ($\sim 140$
km\,s$^{-1}$), and a tendency to show strong brightness variations
($\sim$1 mag over 20 years). For the last $\sim$7 years it has been
exhibiting a continuous optical and near-IR brightening along with a
change of the emission-line spectrum appearance and cooling of the
star's photosphere. We present the object's $V$--band light curve,
analyze the spectral variations, and compare the observed properties
with those of other recognized Galactic LBVs, such as AG\,Car and
HR\,Car. Overall we conclude the MWC\,930 is a bona fide Galactic
LBV that is currently in the middle of an S\,Dor cycle.

\section{Introduction}

Luminous Blue Variables are evolved massive stars that undergo an
evolutionary phase associated with a very strong mass loss. The
phase seems to have a short duration [1] that, in combination with a
high mass, makes the population of LBVs very small (see [2] for a
recent census of Galactic LBVs). Unlike fast winds of most
supergiants with terminal velocities of thousands km\,s$^{-1}$,
winds of LBVs are typically slow and dense. These properties in some
cases are responsible for formation of circumstellar (CS) dust
around LBVs. Other features of LBVs include eruptions on time scales
of years when an optical brightening is accompanied by a
photospheric cooling and corresponding changes in the ionization
state of the CS gas. These variations, which are also called S\,Dor
cycles, along with the wind signatures (narrow spectral lines and
rich emission-line spectra) and B--A spectral types allow to
identify LBVs almost unambiguously. They are also known to undergo
giant eruptions (e.g., P\,Cyg near 1600 and $\eta$\,Car in 1830's),
but these events are rare and much harder to catch.

In our previous paper [3] we suggested that the emission-line star
MWC\,930 = V446\,Sct is an LBV candidate. Our spectra of the object
taken between 2000 and 2004 showed strong and narrow Balmer lines in
addition to those of He {\sc i} and numerous Fe {\sc ii} lines. They
also showed absorption lines typically present in luminous stars
(e.g., Ne {\sc i}, N {\sc ii}, and Al {\sc iii}). Combining our
photometric data obtained in 1980's and 1990's [4] with those of the
ASAS--3 survey (since 1999, [5]), we found that MWC\,930 brightened
by $\sim$0.5 mag in the $V$--band in 2004--2005. We followed the
object's brightness and spectral variations since that time and now
have a firm evidence that the mentioned brightening was an initial
stage of a typical S\,Dor cycle [1].

We describe our observations in Section \ref{observations}, the
observational results are presented in Section \ref{results},
modeling of the spectrum is presented in Section \ref{modeling}, the
observed behavior of the star and its derived parameters are
compared those of other LBVs in Section \ref{discussion}, and
conclusions are made in Section \ref{conclusions}.

\section{Observations}\label{observations}

Thirty three photometric $BVRI$ ($RI$ of the Cousins photometric
system) observations were obtained in September 2006 -- October 2013
with robotic PROMPT telescopes located in Chile [6]. Transformation
equations between the instrumental and standard photometric systems
were derived on several nights by taking images of fields that
contain standard stars from [7]. The data were reduced with the {\it
daophot} package in IRAF.

\begin{figure}[htb]
\begin{center}
\vspace*{-1.0cm} \resizebox{9.0cm}{!}{\includegraphics{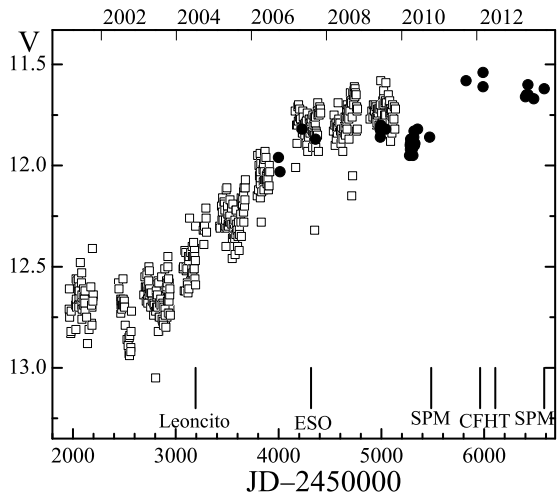}}
\caption{$V$--band light curve of MWC\,930 in 2001--2013. Open
squares show data from ASAS--3, filled circles show our PROMPT data.
Short lines show times of the spectral observations and the
observatory name. The upper horizontal axis tick marks show
beginning time of the indicated years. \label{fig1}}
\end{center}
\end{figure}

Our photometric data were supplemented by $V$--band observations
from the ASAS--3 survey referred above that currently offers
publicly available data until 2011. Individual ASAS--3 datapoints
that typically come from five separate telescopes were averaged.
Since all the stars in our photometric fields are fainter and much
bluer than MWC\,930, we used the overlapping part of the ASAS--3
light curve to calibrate the $V$--band brightness of the object.
Only relative variations were measured for the color-indices. A
light curve of MWC\,930 in the $V$--band is shown in Figure
\ref{fig1}.

Spectroscopic observations were obtained at various observatories in
both hemispheres thanks to the object's nearly equatorial location.
Additionally we retrieved a spectrum of the object obtained at the
ESO with the spectrograph FEROS and reported in [8]. All the spectra
(except the low-resolution one taken at Lick) were obtained with
\'echelle spectrographs. The spectra obtained at the Complejo
Astron\'omico El Leoncito and San Pedro Martir were reduced in a
standard way with the {\it echelle} package in IRAF. Observations
obtained at the CFHT were reduced with the Upena and Libre-ESpRIT
software packages. The Lick spectrum was absolutely calibrated using
a standard star HIP\,96379 (G2 {\sc v}, $V = 8.83$ mag, $K = 7.25$
mag). The observing log is presented in Table \ref{tab1}.

\begin{table}[htb]
  \begin{center}
  \caption{Spectroscopic observations of MWC\,930 in 2004--2013}
  \label{tab1}
  \begin{tabular}{ccrcc} \hline
Date & JD-2450000 & Telescope & $R$ & Sp.range \\
\hline
2004/07/06$^a$ & 3193.674 & 2.1\,m Leoncito & 15000 & 4320--6830\\
2007/08/04 & 4316.739 & 2.2\,m ESO      & 48000 & 4200--8850\\
2010/10/15 & 5485.000 & 2.1\,m SPM      & 15000 & 4000--6830\\
2012/02/03 & 5960.669 & 3.6\,m CFHT     & 65000 & 3600--10500\\
2012/06/29 & 6107.521 & 3.6\,m CFHT     & 65000 & 3600--10500\\
2013/07/21 & 6494.898 & 3.0\,m Lick     & 700   & 4600--23500\\
2013/10/17 & 6583.585 & 2.1\,m SPM      & 15000 & 4556--8130\\
\hline
\end{tabular}
\begin{list}{}
\item Telescopes and spectrographs used: REOSC at the 2.1\,m of the Complejo Astron\'omico El
Leoncito (Argentina), FEROS at the 2.2\,m ESO/MPG of the La Silla
Observatory (Chile), REOSC at the 2.1\,m of the San Pedro Martir
(SPM) Observatory (Mexico), ESPaDOns at the 3.6\,m CFHT (Hawaii,
USA), and NIRI at the 3\,m Shane telescope at the Lick Observatory
(California, USA).
\item $^a$ -- the spectrum was reported in [3] and is mentioned here
for comparison with the visual maximum data
\end{list}
\end{center}
\end{table}

\section{Results}\label{results}

In our previous paper [3] we reported that MWC\,930 exhibited a
stable optical brightness at $V \sim 12.7\pm$0.2 mag in 1989--2003
(we call this period the visual minimum). Since that time the object
got $\sim$ 1.2 mag brighter (Fig.\,\ref{fig1}). The current
brightness level with $V \sim 11.7\pm$0.2 mag (since 2007) is called
the visual maximum.

Color-indices of MWC\,930 changed only slightly during the
brightening. In particular, $B-V$ became 0.10$\pm$0.05 mag larger,
while $V-R$ and $R-I$ remained virtually the same (with a typical
scatter of 0.1 mag). The near-IR magnitudes were determined by
integration of the absolute fluxes in our Lick spectrum ($J$ = 5.7
mag, $H$ = 4.9 mag, $K$ = 4.4 mag, with an uncertainty of $\sim$0.1
mag in all three bands). All the magnitudes are $\sim$1 mag brighter
than those in the visual minimum [4].

Additionally, MWC\,930 was detected in the 9--$\mu$m band by the
{\it AKARI} all sky survey [9] with a flux of 3.25$\pm$0.02 Jy. This
is $\sim$ 0.6 mag brighter than its MSX flux measured during the
visual minimum state [10]. Taking into account that the {\it AKARI}
data were taken in the beginning of the outburst (in 2006--2007) and
the object got brighter since then, the relative fluxes of MWC\,930
in the range 0.4--9 $\mu$m changed very little between the visual
minimum and maximum.

\begin{figure}[htb]
\begin{center}
\resizebox{14.0cm}{!}{\includegraphics{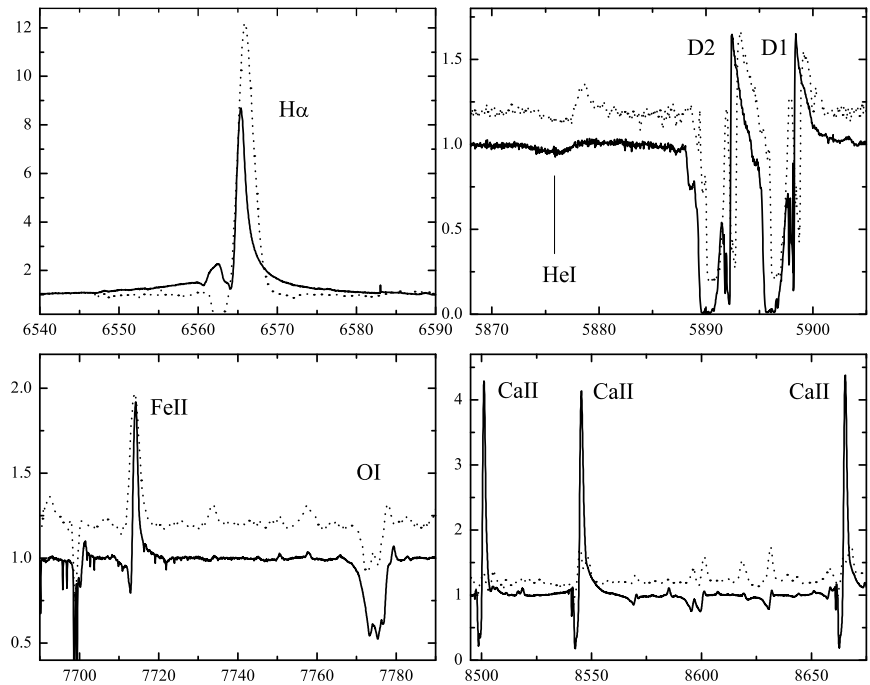}} \vspace*{-1.0cm}
\caption{Spectra of MWC\,930 obtained at the visual minimum (dotted
lines) and maximum (solid lines). The visual minimum spectra were
obtained in October 2001 (upper panels) and July 2004 (lower
panels). The visual maximum spectrum was obtained in June 2012 (all
panels). The intensity is shown normalized to the underlying
continuum, and the wavelengths are in Angstroms and
heliocentric.\label{fig2}}
\end{center}
\end{figure}

The high-resolution spectra obtained during the visual maximum show
that the spectrum has changed dramatically compared to that observed
during the visual minimum. Balmer lines became somewhat weaker, and
the blueshifted emission peak in the H$\alpha$ line became more
pronounced (Beals type {\sc iii}, [11]). Very weak absorption
components of the Paschen lines became much stronger, while their
emission components got very weak. The number of emission lines
increased significantly, while most absorption lines detected
earlier became weaker or disappeared. Most weak metallic lines have
P\,Cyg type profiles (a similar description was presented in [12]
with no details). The IR calcium triplet lines (8498, 8542, and 8662
\AA) show very strong P\,Cyg type profiles reaching $\sim$4
continuum intensities at the peak. The IR oxygen triplet (7772--7775
\AA) in absorption, which was present during the visual minimum but
not mentioned in [3,8], became stronger. Changes in several regions
of the spectrum of MWC\,930 between the visual minimum (2001--2004)
and maximum (2012) are presented in Fig.\,\ref{fig2}.
\vspace*{-1.0cm}

\begin{figure}[htb]
\begin{center}
\resizebox{14.0cm}{!}{\includegraphics{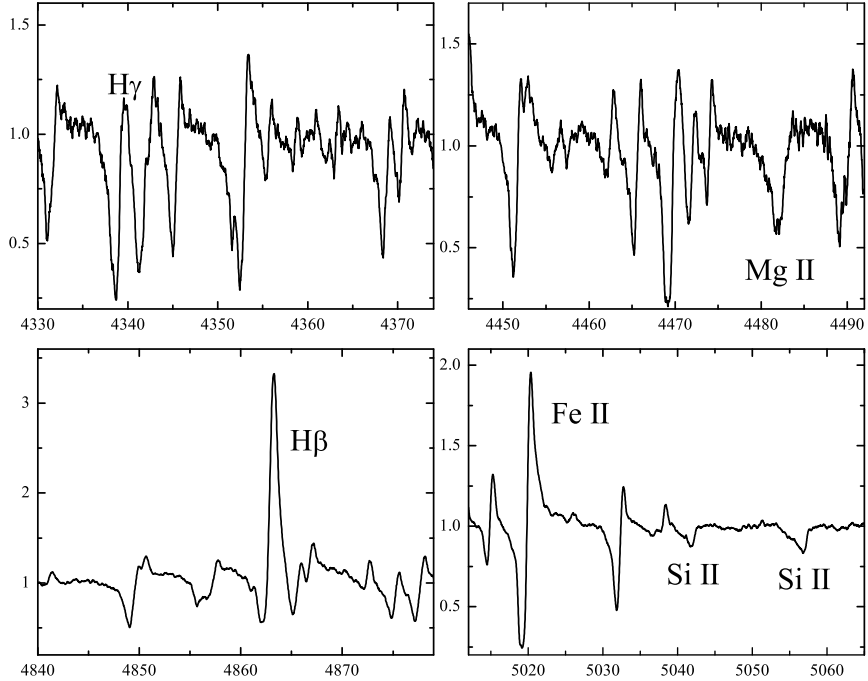}} \vspace*{-1.0cm}
\caption{Parts of the CFHT spectrum of MWC\,930 obtained on 2012
June 29. The intensity and wavelengths are in the same units as in
Fig.\,\ref{fig2}.\label{fig3}}
\end{center}
\end{figure}

MWC\,930 is a relatively optically faint star that also has a very
red color ($B-V \sim$ 2.5 mag). Therefore, long exposure times are
required even with 2--3\,m class telescopes to get a high quality
spectrum in the blue wavelength range. Our spectrum obtained at CFHT
in June 2012 has a signal-to-noise ratio of $\ge$ 30 in the
continuum between 4300 and 5000 \AA\ (and gets to over 100 near
H$\alpha$) that allows us to study lines in this spectral region.
Parts of it are shown in Fig.\,\ref{fig3} for the first time.

All the spectra taken after the beginning of the brightening are
very similar to each other. This is not surprising, as they refer to
nearly the same brightness ($V$ = 11.6--11.8 mag). Some of their
features allow us to evaluate changes in the star's fundamental
parameters compared to those at the visual minimum (spectral type
B1, T$_{\rm eff} = 22000\pm$5000 K, $\log$ L/L$_{\odot} =
5.5\pm0.2$, [3]).

The spectral type during the visual minimum was mainly constrained
by the presence of He {\sc i} lines in emission. These lines are
weak and pure in absorption at the visual maximum, indicating a
lower effective temperature of the star (see Fig.\,\ref{fig2}).
Another feature is a strong absorption line of Mg {\sc ii} at
4482\AA\ (Fig.\,\ref{fig3}). It was reported in [8] as the basis for
the object's spectral classification as B5--B9. The FEROS spectrum
used in [8] is very noisy in this wavelength region (signal-to-noise
S/N $\sim$9 in the continuum) making this conclusion uncertain. Our
CFHT spectrum taken in June 2012 has a S/N $\sim$30 near the
magnesium line and confirms that it is indeed strong and has an
equivalent width (EW) of 0.95 \AA. It is even stronger than that in
spectra of normal supergiants (with no line emission), in which it
increases as the effective temperature drops and reaches $\sim$0.7
\AA\ at the spectral type F0 (e.g., [13]). This implies that the
line is affected by the stellar wind as suggested in [8]. In this
case the line EW is not a good indicator of the star's spectral
type.

\begin{figure}[htb]
\begin{center}
\vspace*{-1.0cm} \resizebox{12.0cm}{!}{\includegraphics{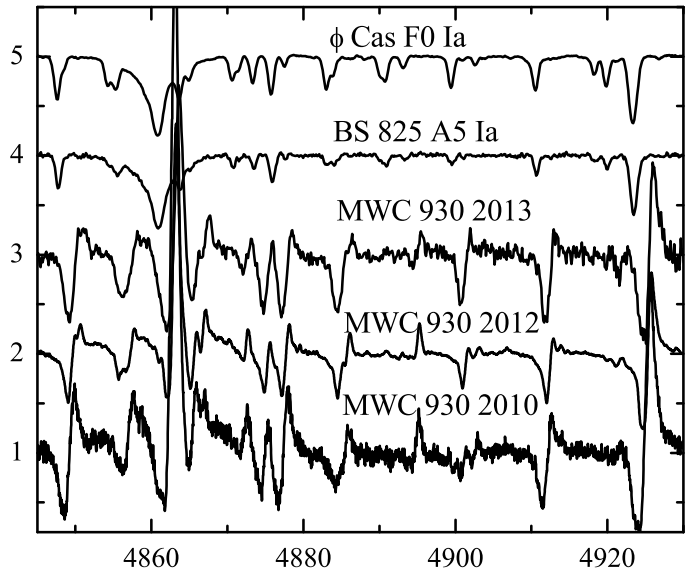}}
\caption{Evolution of the spectrum of MWC\,930 in 2010--2013 and
comparison with spectra of normal stars. The spectra of BS\,825 and
$\phi$ Cas were obtained with a resolving power of $\sim$10000 at
the Three College Observatory of the University of North Carolina at
Greensboro. The intensity and wavelengths are in the same units as
in Fig.\,\ref{fig2}.\label{fig4}}
\end{center}
\end{figure}

Nevertheless, we can estimate how cool the photosphere of MWC\,930
becomes at the visual maximum. The strength of the emission lines
decreases and absorption lines of singly ionized metals get stronger
from 2010 to 2013. Comparison with spectra of normal supergiants
shows that the absorption-line spectrum resembles those of A5--F0
supergiants (see Fig.\,\ref{fig4}). Therefore, the star's effective
temperature is around 8000 K.

As the star got cooler, such luminosity indicators as the Si {\sc
iii} 5739 \AA\ line [14] that appears in the spectra of early
B--type supergiants became unavailable. However, other luminosity
criteria can be applied. The EW of the IR oxygen triplet at
7772--7775 \AA\ has been shown to be a reasonably good luminosity
calibrator in B--F type supergiants (e.g., [15]). The average EW of
the triplet in our spectra is 2.7$\pm$0.1 \AA\ that corresponds to
an absolute visual magnitude of M$_{V} \sim -9.0$ mag. The triplet
shows a weak emission component on the red side (Fig.\,\ref{fig2}).
Therefore, its properties may be affected by the CS gas and have a
limited use. Nevertheless, the triplet EW is very similar (2.8 \AA)
to that of the A2--hypergiant IRC+10420 (V1302\,Aql) which has a
bolometric absolute magnitude of M$_{\rm bol} \sim -9.5$ mag [16].
Both estimates are within the uncertainty from our earlier result
for the object's luminosity ($\log$ L/L$_{\odot} = 5.5\pm0.2$, [3]).

Most spectral lines show changes of positions and intensity with
time. This is most likely due to a variable mass loss, whose
strength affects the density and ionization structure of the stellar
wind. Table\,\ref{tab2} shows the variability of the H$\alpha$ line
profile since our last spectroscopic observation during the visual
minimum. The line blue peak and central depression that form in
front of the star in an optically-thick part of the wind show
significant variations. The red peak that forms in the receding part
of the wind shows a stable position, but a variable strength which
decreases with time. The latter may be due to a lowering star's
effective temperature and a smaller ionized part of the wind.

\begin{table}[htb]
  \begin{center}
  \caption{Parameters of the H$\alpha$ line in the spectrum of MWC\,930 in
  2004--2013} \vskip 0.2cm
  \label{tab2}
  \begin{tabular}{lrcrccrc}
\hline\noalign{\smallskip}
Date & \multicolumn{2}{c}{Blue} & \multicolumn{2}{c}{Center} & \multicolumn{2}{c}{Red}& EW,\AA \\
\cline{2-3}\cline{4-5}\cline{6-7}
               & Vr    & I  & Vr & I & Vr & I & \\
\noalign{\smallskip}\hline\noalign{\smallskip}
2004/07/06$^a$ & $-$94 & 2.0&$-$29& 0.48&116 &18.9& 54\\
2007/08/04     & $-$64 & 2.2& 23  & 0.73&114 &14.5& 39\\
2010/10/15     & $-$6  & 3.7& 49  & 1.83&118 &12.8& 36\\
2012/02/03     & $+$35 & 2.4& 60  & 1.37&123 & 8.4& 19\\
2012/06/29     & $-$17 & 2.2& 58  & 1.20&118 & 8.8& 19\\
2013/10/17$^b$ & $-$63 & 1.7& 16  & 0.69&116 & 8.3& 19\\
\noalign{\smallskip}\hline
\smallskip
\end{tabular}
\begin{list}{}
\item Date of observation is listed in column 1; the radial velocity
in km\,s$^{-1}$ and the intensity in units of the nearby continuum
of the H$\alpha$ line profile parts are listed in columns 2--3 (blue
emission peak), 4--5 (central depression), and 6--7 (red emission
peak), respectively; column 8 lists the EW of the strongest (red)
line peak.
\item $^a$ -- the spectrum was reported in [3]
\item $^b$ -- the H$\alpha$ line in this spectrum has an additional absorption component
at $-$129 km\,s$^{-1}$ and 1.11 continuum intensity.
\end{list}
\end{center}
\end{table}

\begin{figure}[htb]
\begin{center}
\begin{tabular}{cc}
\resizebox{8.7cm}{!}{\includegraphics{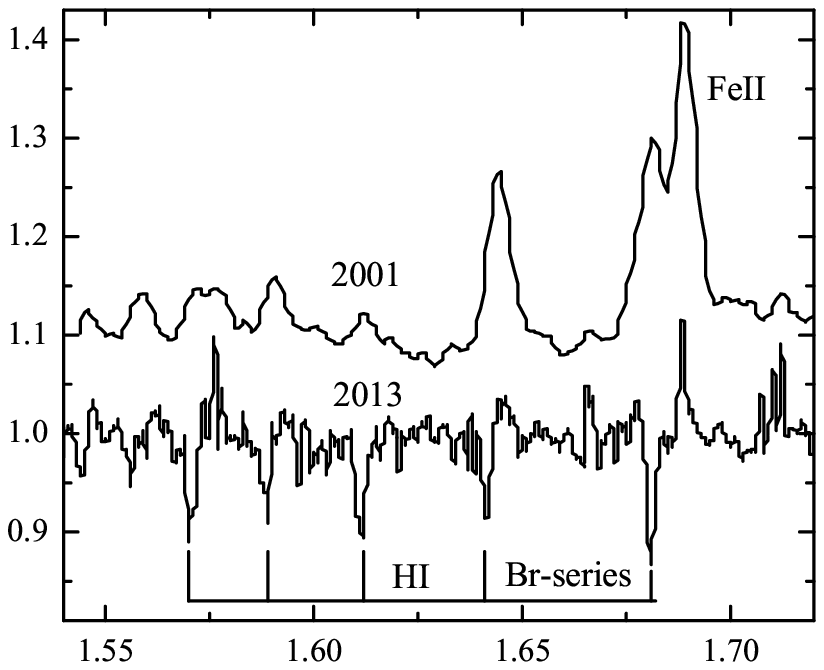}}&
\hspace*{-1.0cm} \resizebox{8.7cm}{!}{\includegraphics{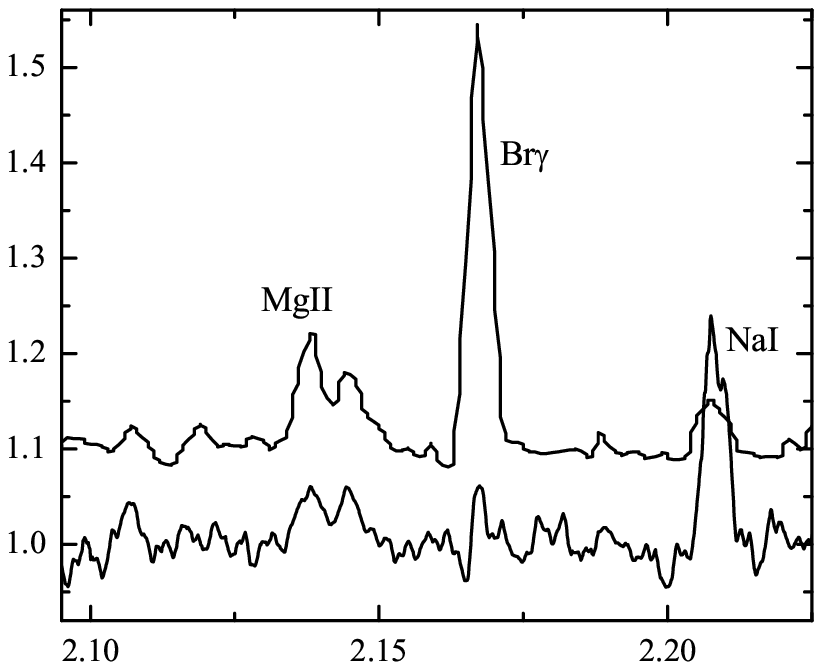}}\\
\end{tabular}
\vspace*{-1.0cm} \caption{Parts of the Lick near-IR spectrum of
MWC\,930 obtained on 2013 July 21. The intensity is normalized to
the underlying continuum, and the wavelengths are in $\mu$m. The
spectrum from 2001 is shifted up by 0.2 continuum intensity.
\label{fig5}}
\end{center}
\end{figure}

The near-IR spectrum of MWC\,930 (see Fig.\,\ref{fig5}) shows
significant changes compared to our spectrum obtained in 2001 [3].
The Brackett series hydrogen lines turned into absorption except for
Br$_{\gamma}$. At the same time, a line of neutral sodium at 2.22
$\mu$m became stronger. This is also consistent with a lower
effective temperature and a smaller ionized part of the wind.

\section{Modeling}\label{modeling}

In order to derive stellar wind properties and verify the results of
our analysis of the observational data, we used a grid of
pre­calculated models of the radiation transfer code CMFGEN. This
code [17] allows for the calculation of non­-LTE level populations
of the ionization states for all included elements, as well as
provides the emergent spectra of the star. It has been used to model
spectra of LBVs, O­-type and Wolf--Rayett stars [e.g., 18].

\begin{figure}[htb]
\begin{center}
\resizebox{14.0cm}{!}{\includegraphics{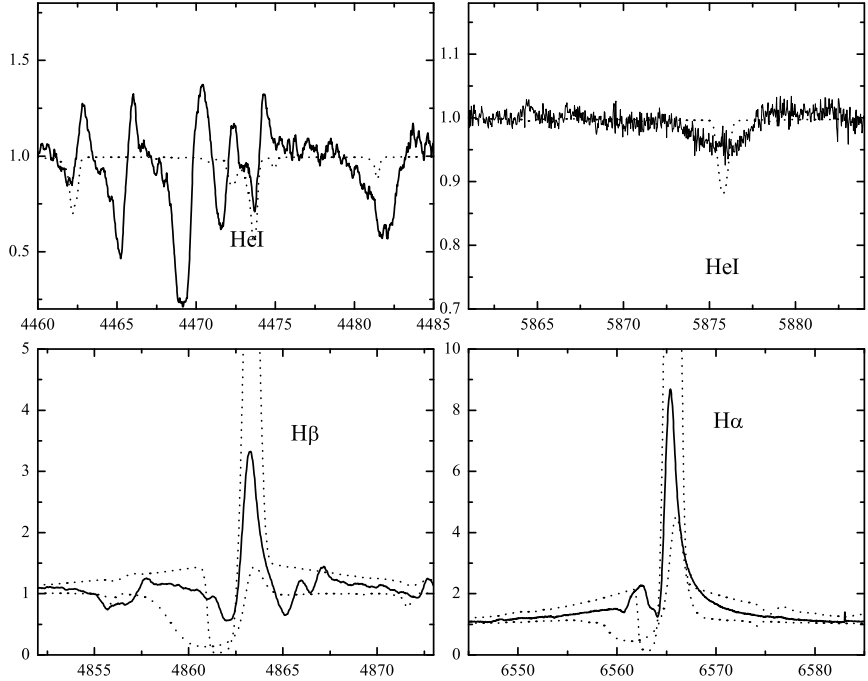}} \vspace*{-1.0cm}
\caption{Modeling of selected optical He {\sc i} lines at 4471 and
5876 \AA\, (upper row), and the H$\alpha$ and H$\beta$ lines (bottom
row) for MWC\,930. Parts of our CFHT spectrum taken on 2012 June 29
are shown by solid lines, while parts of the model spectra are shown
by dotted lines. One model spectrum with a mass loss rate of \.M =
3\,10$^{-5}$ M$_{\odot}$\,yr$^{-1}$ is shown for the He {\sc i}
lines. Two model spectra with mass loss rates of \.M = 3\,10$^{-5}$
M$_{\odot}$\,yr$^{-1}$ (with a weaker line emission) and \.M =
5.6\,10$^{-5}$ M$_{\odot}$\,yr$^{-1}$ are shown for the Balmer
lines. The intensity and wavelengths are in the same units as in
Fig.\,\ref{fig2}.\label{fig6}}
\end{center}
\end{figure}

Figure\,\ref{fig6} shows two fits to a selection of Balmer and He
{\sc i} lines that are reasonably close to the observed line
profiles. The weak or non­existent He {\sc i} lines indicate a low
effective temperature (T$_{\rm eff} \le$ 10000 K), and the narrow
H$\alpha$ and H$\beta$ suggest a low surface gravity of $\log$ g
$\le$ 1. It is also obvious from Fig.\,\ref{fig6} that the terminal
velocity is low (v$_{\infty} \le$ 100 km\,s$^{-1}$). Based on these
initial constraints, both models were calculated for T$_{\rm eff}$ =
8233 K and $\log$ g = 0.9. The spherical stellar wind was described
by a $\beta$ velocity law with $\beta$ = 3 and a terminal velocity
of 73 km\,s$^{-1}$ (see [18] for a description of the modeling
features). This is consistent with our earlier estimate for the
visual minimum state [3].

The mass loss rate was set to be \.M = 3\,10$^{-5}$
M$_{\odot}$\,yr$^{-1}$ and \.M = 5.6\,10$^{-5}$
M$_{\odot}$\,yr$^{-1}$. The lower mass loss rate model produces a
too weak emission in the H$\alpha$ and H$\beta$, and the higher mass
loss rate model give a too strong emission in these lines.

Since our grid was originally created for modeling AG\,Car
(including elemental abundances adopted from [18]) which has a
higher mass loss rate than MWC\,930, we can only roughly constrain
the mass loss at \.M = $(4\pm1)\,10^{-5}$ M$_{\odot}$\,yr$^{-1}$.
This estimate is over an order of magnitude larger than the one
derived for the visual minimum (1.5\,10$^{-6}$
M$_{\odot}$\,yr$^{-1}$). Such an increase in the mass loss rate
during the visual maximum of MWC\,930 is larger than that (about a
factor of 5) derived for AG\,Car in [18]. This difference may be due
to a lower luminosity of MWC\,930. Nevertheless, the effective
temperatures of both objects at visual maximum are very close to
each other.

\section{Discussion}\label{discussion}

As we showed above, MWC\,930 simultaneously became optically
brighter and cooler. This is a typical behavior of an LBV during an
S\,Dor cycle [1]. Therefore the observational data reported here
prove our previous suggestion about the nature of this object [3].

The observed behavior can be compared to that of other known LBVs.
For example, an S\,Dor cycle of AG\,Car that occurred during the
1990's was well documented in [19]. This object began the eruption
with a brightness of $V \sim$ 8.0 mag in mid--1990 and reached a
maximum brightness of $V =5.6$ mag by the end of 1994, then it
gradually dimmed to $V \sim 7.3$ mag by the end of 1998. As it was
brightening, its spectrum showed signs of lowering excitation. The
following changes were observed in the line profiles between the
beginning of the cycle and a brightness level of $V \sim 6.7$ mag.
Lines of neutral helium, which started from P\,Cyg type profiles
with a strong emission component, have disappeared. The Mg {\sc ii}
4482 \AA\ line turned from a weak emission into a strong absorption.
Lines of singly ionized iron and titanium had the strongest P Cyg
type profiles when the objects was brighter than $V \sim $ 6.7 mag.
Qualitatively all these variations were observed during the
brightening of MWC\,930. The H$\alpha$ profiles of both AG\,Car and
MWC\,930 have a similar structure with a blue peak much weaker than
the red one. These similarities strongly suggest that the processes
in both objects during the described periods were alike.

While most spectral lines (e.g., Fe {\sc ii}, Ti {\sc ii}) in the
spectrum of MWC\,930 exhibit pure P\,Cyg type profiles, the
H$\alpha$ line profile can be described as a combination of a P\,Cyg
type and a double-peaked profile. Such a complicated profile is hard
to explain by a spherically-symmetric wind [19]. It needs either an
additional disk-like wind component or can be due to a bipolar
outflow. Nevertheless, fitting the line's red peak can give an idea
about changes of the mass loss rate (see [19] for AG Car).
Qualitatively, it appears that the wind becomes stronger with time
as the star gets cooler and larger. This process seems to have
continued in 2012--2013, as both emission peaks of the H$\alpha$
line have been getting weaker while the central depression has been
getting deeper (see Table\,\ref{tab2}). The H$\alpha$ emission peaks
are also getting closer together during the visual maximum which is
probably an increasing wind optical depth effect.

Changes in the LBVs' effective temperature and radius during an
S\,Dor cycle are typically calculated assuming a constant bolometric
luminosity [e.g., 19]. This assumption applied to MWC\,930 (which is
qualitatively supported by the above mentioned luminosity estimates)
leads to the following results. Since the object became $\sim$1.2
mag brighter in the $V$--band (in 2012--2013), then the bolometric
correction has got lower by this value. We also need to take into
account the CS contribution to the visual continuum which was
determined to be $\Delta V$ = 0.5 mag in [3]. We neglect this
contribution at the visual maximum, because the ionizing radiation
should be much weaker due to a lower effective temperature of the
star. If we take T$_{\rm eff}$ = 22000 K at the visual minimum and a
bolometric correction of $-2.2$ mag [3], then T$_{\rm eff}$ is
$\sim$ 9000 K for 2012--2013. If T$_{\rm eff}$ = 17000 K (the lowest
value from [3]) that corresponds to a bolometric correction of
$-1.5$ mag [20], then the current T$_{\rm eff} \sim$ 8000 K.
Therefore, we can conclude that MWC\,930 has been keeping roughly
the same bolometric luminosity during the transition between the
visual minimum and maximum.

\begin{figure}[htb]
\begin{center}
\resizebox{10.0cm}{!}{\includegraphics{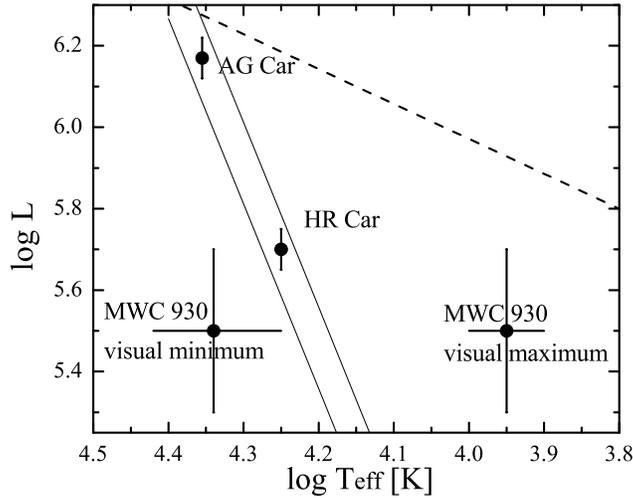}} \caption{A
Hertzsprung-Russell diagram with positions of some LBVs. The solid
lines show the boundaries of the visual minimum brightness
instability strip for LBVs at which the stellar rotation velocity
equals the break-up velocity [20]. The dashed line represents the
Humphreys--Davidson stability limit [1]. Fundamental parameters of
AG\,Car and HR\,Car at a visual minimum are taken from [20].
Luminosity is plotted in the solar units, and temperature in
Kelvins. \label{fig7}}
\end{center}
\end{figure}

With these results for the fundamental parameters of MWC\,930, one
can compare them with those of other LBVs. It has recently been
suggested [21] that LBVs which exhibit strong S\,Dor cycles are fast
rotators, and their parameters are constrained by a critical
rotation limit. We show the positions of MWC\,930 in both brightness
states along with those (in a visual minimum state only) of two
other strongly variable LBVs, AG\,Car and HR\,Car, on the
Hertzsprung-Russell diagram in Fig.\,\ref{fig7}.

The position of MWC\,930 during the visual minimum lies in the
forbidden area (left of the solid line in Fig.\,\ref{fig7}) unless
we assume the lowest star's temperature estimated in [3]. The visual
maximum position violates no current stability limits. As far as the
star's rotational velocity is concerned, we have suspected in [3]
that the broad (and sometimes split) photospheric lines were either
due to a fast rotation or binarity. The lower luminosity of MWC\,930
compared to those of both AG\,Car and HR\,Car is consistent with its
weaker emission-line spectrum and therefore a lower mass loss rate.

As we discussed in [3], the presence of a far-IR excess in the
spectral energy distribution of MWC\,930 was a hint towards the LBV
evolutionary status. Recently an extended shell around the object
was discovered in IR images taken by the Spitzer Space Observatory
and the WISE mission at wavelengths longer than 5 $\mu$m [22]. This
finding is consistent with the idea of a past giant eruption of
MWC\,930. Additionally, analysis of the IR spectrum of MWC\,930
presented in [22] showed that the silicate absorption feature at
$\lambda = 10 \mu$m is due to the interstellar rather than to the CS
extinction. This result indicates a large distance towards the
object and thus supports the high luminosity suggested in [3] and in
this paper.

\section{Conclusions}\label{conclusions}

We reported multicolor optical photometry, high-resolution optical
spectroscopy, and low-resolution flux-calibrated near-IR
spectroscopy of the LBV candidate MWC\,930 obtained in 2006--2013.
The observational data clearly show that the object is undergoing an
S\,Dor cycle. During this period the optical brightness increased by
$\sim$1.2 mag and the spectrum became less excited due to lowering
the star's effective temperature. The spectral features observed at
the visual maximum suggest that T$_{\rm eff}$ became
$\sim$8000--9000 K that is consistent with an assumption of the
bolometric luminosity constancy during the eruption, as is typically
observed in LBVs. Comparison of the spectral behavior and
fundamental parameters of MWC\,930 with other LBVs that exhibit S
Dor cycles (AG\,Car and HR\,Car) suggests that the star's effective
temperature during the visual minimum was on the lowest side of our
previous estimate ($\sim$17000 K, [3]), while the bolometric
luminosity ($\log$ L/L$_{\odot} \sim$ 5.5, [3]) did not seem to
change.

\vspace*{0.5cm}

\section{Acknowledgements}

We are grateful to M. Borges Fernandes for retrieving and reducing
the FEROS spectrum of MWC\,930. A.~M. and S.Z. acknowledge support
from DGAPA/PAPIIT project IN100614. This paper is partially based on
observations obtained at the Canada-France-Hawaii Telescope (CFHT)
which is operated by the National Research Council of Canada, the
Institut National des Sciences de l$^{\prime}$Univers of the Centre
National de la Recherche Scientifique de France, and the University
of Hawaii, the 2.2m MPG telescope operated at ESO/La Silla under
program IDs 086.A--9019 and 087.A--9005, the 2.1m telescope of the
San Pedro Martir Observatory, the 2.1m telescope of the Complejo
Astron\'omico El Leoncito, and the PROMPT robotic telescopes located
in Chile and operated by University of North Carolina at Chapel
Hill.

\vspace*{0.5cm}

{\bf References:} \vspace*{0.3cm}

1. R.~M.~Humphreys, K.~Davidson, ``The luminous blue variables:
Astrophysical geysers'', {\it Publications of the Astronomical
Society of the Pacific}, vol. 106, pp. 1025--1051, 1994.

2. J.~S.~Clark, V.~.M.~Larionov, A.~A.~Arkharov, ``On the population
of galactic Luminous Blue Variables'', {\it Astronomy and
Astrophysics}, vol. 435, pp. 239--246, 2005.

3. A.~S.~Miroshnichenko, K.~S.~Bjorkman, M.~Grosso, et al., ``MWC
930 - a new luminous blue variable candidate'', {\it Monthly Notices
of the Royal Astronomical Society}, vol. 364, pp. 335--343, 2005.

4. Yu.~K.~Bergner, A.~S.~Miroshnichenko, R.~V.~Yudin, et al.,
``Observations of emission-line stars with IR excesses. II.
Multicolor photometry of B[e] stars.'', {\it Astronomy and
Astrophysics Supplement Series}, vol. 112, pp. 221--228, 1995.

5. G.~Pojmanski, ``The All Sky Automated Survey. Catalog of Variable
Stars. I. 0 h -- 6 h Quarter of the Southern Hemisphere'', {\it Acta
Astronomica}, vol. 52, pp. 397--427, 2002.

6. D.~E.~Reichart, M.~Nysewander, J.~Moran, et al., ``PROMPT:
Panchromatic Robotic Optical Monitoring and Polarimetry
Telescopes'', {\it Nuovo Cimento C}, vol. 28, pp. 767--770, 2005.

7. A.~U.~Landolt, ``UBVRI photometric standard stars around the
celestial equator'', {\it Astronomical Journal}, vol. 88, pp.
439--460, 1983.

8. A.~Carmona, M.~E.~van~den~Ancker, M.~Audard, et al., ``New Herbig
Ae/Be stars confirmed via high-resolution optical spectroscopy'',
{\it Astronomy and Astrophysics}, vol. 517, A67, 22 pp., 2010.

9. D.~Ishihara, T.~Onaka, H.~Kataza, et al., ``The AKARI/IRC
mid-infrared all-sky survey'', {\it Astronomy and Astrophysics},
vol. 514, A1, 14 pp., 2010.

10. M.~P.~Egan, S.~D.~Price, K.~E.~Kraemer, et al., ``The Midcourse
Space Experiment Point Source Catalog Version 2.3 (October 2003)'',
AFRL--VS--TR--2003--1589, 2003.

11. C.~S.~Beals, ``The Spectra of the P Cygni Stars'', {\it
Publications of the Dominion Astrophysical Observatory}, vol. 9, pp.
1--49, 1953.

12. A.~Lobel, J.~H.~Groh, K.~Torres, N.~Gorlova, ``Long-term
spectroscopic monitoring of LBVs and LBV candidates'', in {\it
Active OB stars: structure, evolution, mass loss, and critical
limits}, Proceedings of the Symposium, Vol. 272, pp. 519--520, 2011.

13. D.~J.~Lennon, P.~L.~Dufton, A.~Fitzsimmons, ``Galactic
B--supergiants. II. Line strengths in the visible -- Evidence for
evolutionary effects?'', {\it Astronomy and Astrophysics Suppl.
Ser.}, vol. 97, pp. 559--585, 1993.

14. A.~S.~Miroshnichenko, H.~Levato, K.~S.~Bjorkman, et al. 2004.
``Properties of Galactic B[e] supergiants. III. MWC 300'', {\it
Astronomy and Astrophysics}, vol. 417, pp. 731--743, 2004.

15. V.~V.~Kovtyukh, N.~I.~Gorlova, S.~I.~Belik, ``Accurate
luminosities from the oxygen 7771--4 \AA\ triplet and the
fundamental parameters of F--G supergiants'', {\it Monthly Notices
of the Royal Astronomical Society}, vol. 423, pp. 3268--3273, 2012.

16.  V.~G.~Klochkova, M.~V.~Yushkin, E.~L.~Chentsov, V.~E.~Panchuk,
``Evolutionary Changes in the Optical Spectrum of the Peculiar
Supergiant IRC+10420'', {\it Astronomy Reports}, vol. 46, pp.
139--152, 2002.

17. D.~J.~Hillier, D.~L.~Miller, ``The Treatment of Non-LTE Line
Blanketing in Spherically Expanding Outflows'', {\it Astrophysical
Journal}, vol. 496, pp. 407--427, 1998.

18. J.~H.~Groh, D.~J.~Hillier, A.~Damineli, et al., ``On the Nature
of the Prototype Luminous Blue Variable AG\,Carinae. I. Fundamental
Parameters During Visual Minimum Phases and Changes in the
Bolometric Luminosity During the S--Dor Cycle'', {\it Astrophysical
Journal}, vol. 698, pp. 1698--1720, 2009.

19. O.~Stahl, I.~Jankovics, J.~Kov\'acs, et al., ``Long-term
spectroscopic monitoring of the Luminous Blue Variable AG Carinae'',
{\it Astronomy and Astrophysics}, vol. 375, pp. 54--69, 2001.

20. A.~S.~Miroshnichenko, ``New photometric calibration of the
visual surface brightness method'', in {\it Fundamental Stellar
Properties: The Interaction between Observation and Theory}, Proc.
IAU Symp. 189, pp. 50--53, 1997.

21. J.~H.~Groh, A.~Damineli, D.~J.~Hillier, et al., ``Bona Fide,
Strong-Variable Galactic Luminous Blue Variable Stars are Fast
Rotators: Detection of a High Rotational Velocity in HR Carinae'',
{\it Astrophysical Journal}, vol. 705, pp. L25--L30, 2009.

22. L.~Cerrigone, G.~Umana, C.~S.~Buemi, et al., ''Spitzer
observations of a circumstellar nebula around the candidate Luminous
Blue Variable MWC\,930'', {\it Astronomy and Astrophysics}, vol.
562, A93, 9 pp., 2014.

\end{document}